\documentclass[aps,twocolumn,showpacs,preprintnumbers,prb,amsmath,amssymb,amsfonts,superscriptaddress,floatfix]{revtex4-1}
\usepackage{graphicx}
\usepackage{dcolumn}
\usepackage{bm}
\usepackage[version=3]{mhchem} 
\usepackage{color}
\usepackage{array}
\usepackage{multirow}

\begin{document}
\title{Optical and Electronic Properties of Doped $p$-type CuI:
Explanation of Transparent Conductivity from First Principles}

\author{Yuwei Li}

\author{Jifeng Sun}

\author{David J. Singh}
\email{singhdj@missouri.edu}
\affiliation{Department of Physics and Astronomy, University of Missouri, Columbia, MO 65211-7010 USA}
\date{\today}

\begin{abstract}
We report properties of the reported transparent conductor CuI, 
including the effect of 
heavy $p$-type doping. The results, based on first principles calculations,
include
analysis of the electronic structure and calculations of optical and dielectric
properties. We find that the origin of the favorable transparent
conducting behavior lies in the absence in the
visible of strong interband transitions between deeper valence bands
and states at the valence band maximum that become empty with $p$-type
doping. 
Instead, strong interband transitions to the valence band maximum
are concentrated in the infrared with energies below 1.3 eV.
This is contrast to the valence bands of many wide band gap materials.
Turning to the mobility we find that the states at the valence band
maximum are relatively dispersive. This originates from their antibonding
Cu $d$ - I $p$ character.
We find a modest enhancement of the Born effective charges relative to
nominal values, leading to a dielectric constant $\varepsilon(0)$=6.3.
This is sufficiently large to reduce ionized impurity scattering, leading to
the expectation that the properties of CuI can be still can be significantly
improved through sample quality.
\end{abstract}

\maketitle

\section{Introduction}

Transparent conductors (TCs),
such as the transparent conducting oxides (TCOs),
are compounds that combine low visible light absorption
with high electrical conductivity.\cite{TCO1,TCO2}
They are important for opto-electronic devices including solar cells and
displays, as well as for applications such as smart windows.
\cite{TCO3,TCO4,TCO5,TCO6,TCO7}
Remarkably, in spite of industrial interest and extensive research, the
number of established high performance TCs suitable
for applications is relatively small. These include $n$-type oxides based
on In,
especially Sn doped In$_2$O$_3$,
\cite{ITO1,ITO2}
$n$-type stanates including BaSnO$_3$,
\cite{wang,BaSnO3_1,BaSnO3_3,singh-basno3,BaSnO3_4}
ZnO based materials \cite{ZnO} and In-Ga-Zn oxides. \cite{TCO6}

Good $p$-type TCs are less common, but are enabling for some
applications, for example in transparent electronics. However,
these $p$-type materials
generally have lower performance than state-of-the-art
$n$-type TCOs.
This fact has motivated several studies of potential $p$-type TCs,
leading to the discovery of several new $p$-type TC materials.
Known $p$-type TCs include, for example,
Cu$^{+}$ oxides\cite{CuAlO2,duan,LaOCuS,CuCrO2},
Sn$^{2+}$ compounds\cite{Ba/SrSn2O3,SnO_1,SnO_2,SnO_3},
BaSnO$_3$,\cite{pBaSnO3_1,pBaSnO3_2}
and Ba$_2$BiTaO$_6$.\cite{Ba2BiTaO6}

Recently, CuI has been identified as a good
$p$-type TC material.
\cite{CuI_1,CuI_2,CuI_3,CuI_4,CuI_5,CuI_6}
CuI has a cubic structure and
is also compatible with solar cells and amenable to thin film growth,
including growth on glass.
\cite{CuI_3,compa_1,compa_2,compa_3,zhu}
Here we present a first principles study of its properties as
related to TC behavior, especially the issue of simultaneous
conductivity and transparency in a $p$-type material.

We find that a
key aspect is the absence of strong interband transitions in the
visible when doped, combined with the particular bonding of the material,
which favors both $p$-type doping and dispersive valence bands. 
We do not find strongly enhanced Born effective charges that
would lead to unusually high dielectric constants and screening of
ionized impurities, in contrast to some other
high mobility halide semiconductors. \cite{du2010enhanced,ming}
Instead we find a modest enhancement, which nonetheless provides
enough screening to suggest that reported $p$-type CuI are not near
the limit for increase of mobility via sample quality improvements.

\section{Structure and Methods}

CuI has a zinc-blende structure at temperatures below 643 K,
($\gamma$ phase, space group $F\bar{4}$3$m$).\cite{CuI_stru}
The ambient temperature experimental lattice parameter is
$a$ = 6.058 {\AA},\cite{S6stru} and we used this value.
We performed first principles
calculations using the
projector augmented-wave (PAW) method,\cite{PAW} as implemented in
the VASP code\cite{VASP} and the general potential
linearized augmented planewave (LAPW)\cite{LAPW} method as implemented in the
WIEN2k code. \cite{wien2k}
We used these two codes because of the ability to apply
the virtual crystal approximation and carry out accurate optical
calculations provided in WIEN2k and the need to perform hybrid functional
and dielectric calculations, which are more convenient in VASP.

We treated the
3$d^{10}$4$s^1$ shells of Cu and the 5$s^2$5$p^5$ shells of I as
valence electrons with PAW pseudopotentials in our VASP calculations.
These were used
with kinetic energy cutoffs of 300 eV. We tested to ensure that this
was adequate.
The k-point meshes for sampling the Brillouin zone were at a
grid spacing of 2$\pi \times$ 0.02 \AA$^{-1}$ or better, including
the hybrid functional calculations.
Born effective charges were calculated using density functional
perturbation theory as implemented in VASP.
We used sphere radii of 2.46 Bohr for Cu and I and basis set cut-offs,
$k_{max}$, set by the criterion $R_{min}k_{max}$ = 9.0 in the LAPW
calculations; here $R_{min}$ is the LAPW sphere radius of 2.46 Bohr.
Spin-orbit coupling (SOC) was included in all electronic structure and
optical property calculations.
The spin-orbit calculation was performed in WIEN2k using the
second variational approach, \cite{LAPW} in which the
relativistic problem is solved using a basis set consisting of the
scalar relativisitic band states. For this purpose all occupied states,
plus unoccupied states up to 5 Ry were employed.

\begin{figure}[tbp]
\includegraphics[width=0.9\columnwidth]{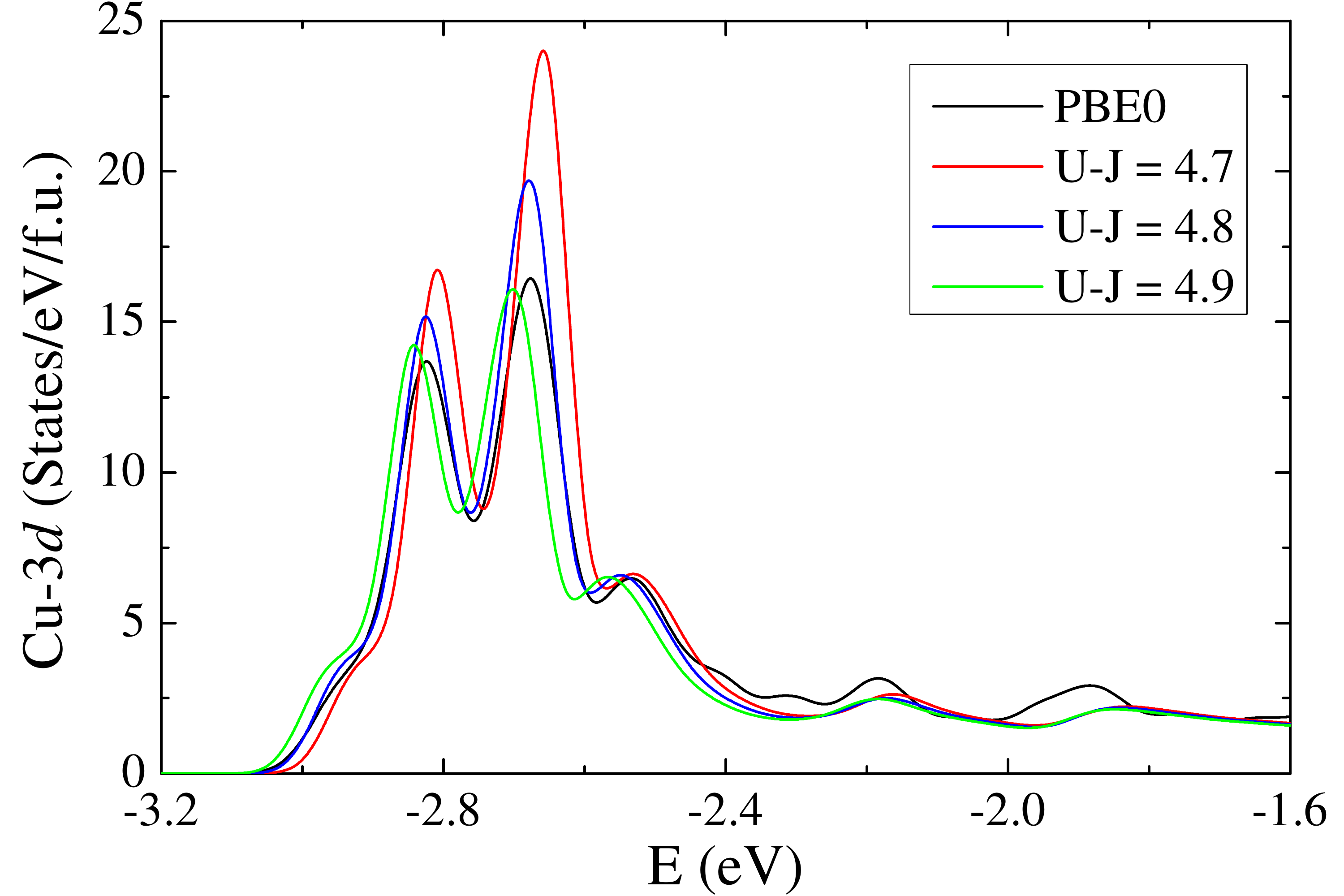}
\caption{Calculated valence band density of states
showing the Cu-3$d$ states with the PBE0 hybrid functional (black)
and GGA+U method when U= 4.7 eV (red), 4.8 eV (blue) and 4.9 eV (green).
Here J is set to 0 and the energy zero is at the valence band maximum.}
\label{Cu-3d}
\end{figure}

Both a correct band gap and a correct position of the Cu $d$ states
in the I $p$ valence bands are important ingredients for obtaining
realistic optical and transport properties.
We used the generalized gradient approximation (GGA) of
Perdew, Burke, and Ernzerhof (PBE), \cite{pbe}
and applied the GGA+U method to describe the Cu $d$ states in
LAPW calculations with the WIEN2k code.
Transport coefficients were obtained using the BoltzTraP code.\cite{boltztrap}
The value of the parameter U was determined based on
hybrid functional calculations with the PBE0 functional.\cite{PBE0}
These PBE0 calculations were done using VASP.
Good agreement between the PBE0 and GGA+U calculations for the
position of the Cu $d$ states was obtained
for U = 4.8 eV, as shown in Fig. \ref{Cu-3d}, which shows the region
of the upper crystal field peak of the Cu $d$ density of states.
This is the energy range that is crucial for the interband optical
absorption in doped $p$-type samples, as discussed below.
Importantly, this choice of U in the GGA+U method also yields very good
agreement for the shape and peak structure of the Cu $d$ density of states.
Besides the good agreement between our HSE0 and GGA+U calculations
for the shape and position of the Cu $d$ states, we note that these
are also in good agreement with results from valence band photoemission
experiments.
\cite{generalov}

\begin{figure}[tbp]
\includegraphics[width=0.9\columnwidth]{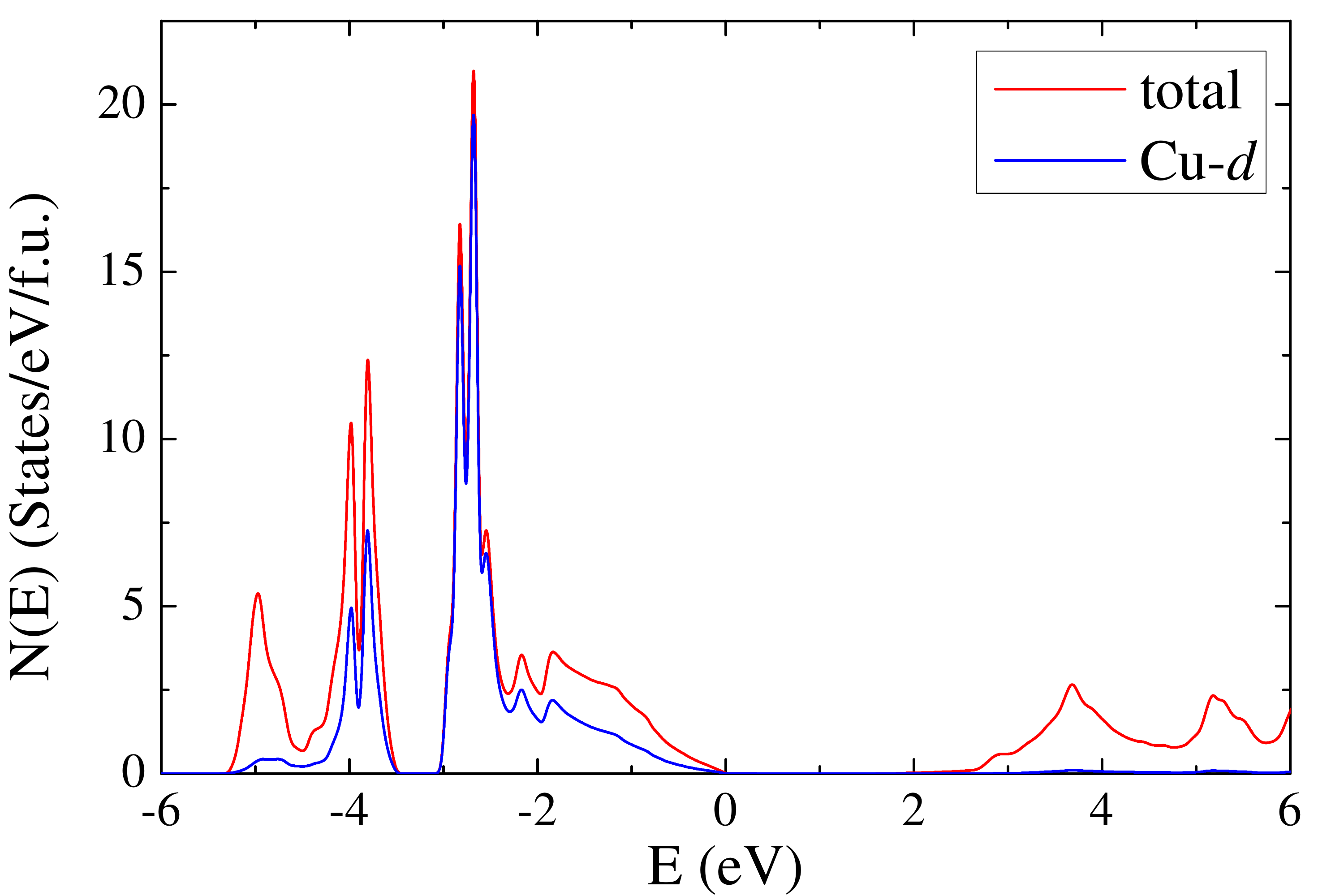}
\caption{GGA+U density of states and Cu $d$ projection without
correction of the band gap.}
\label{dosfull}
\end{figure}

\begin{figure}[tbp]
\includegraphics[width=0.9\columnwidth]{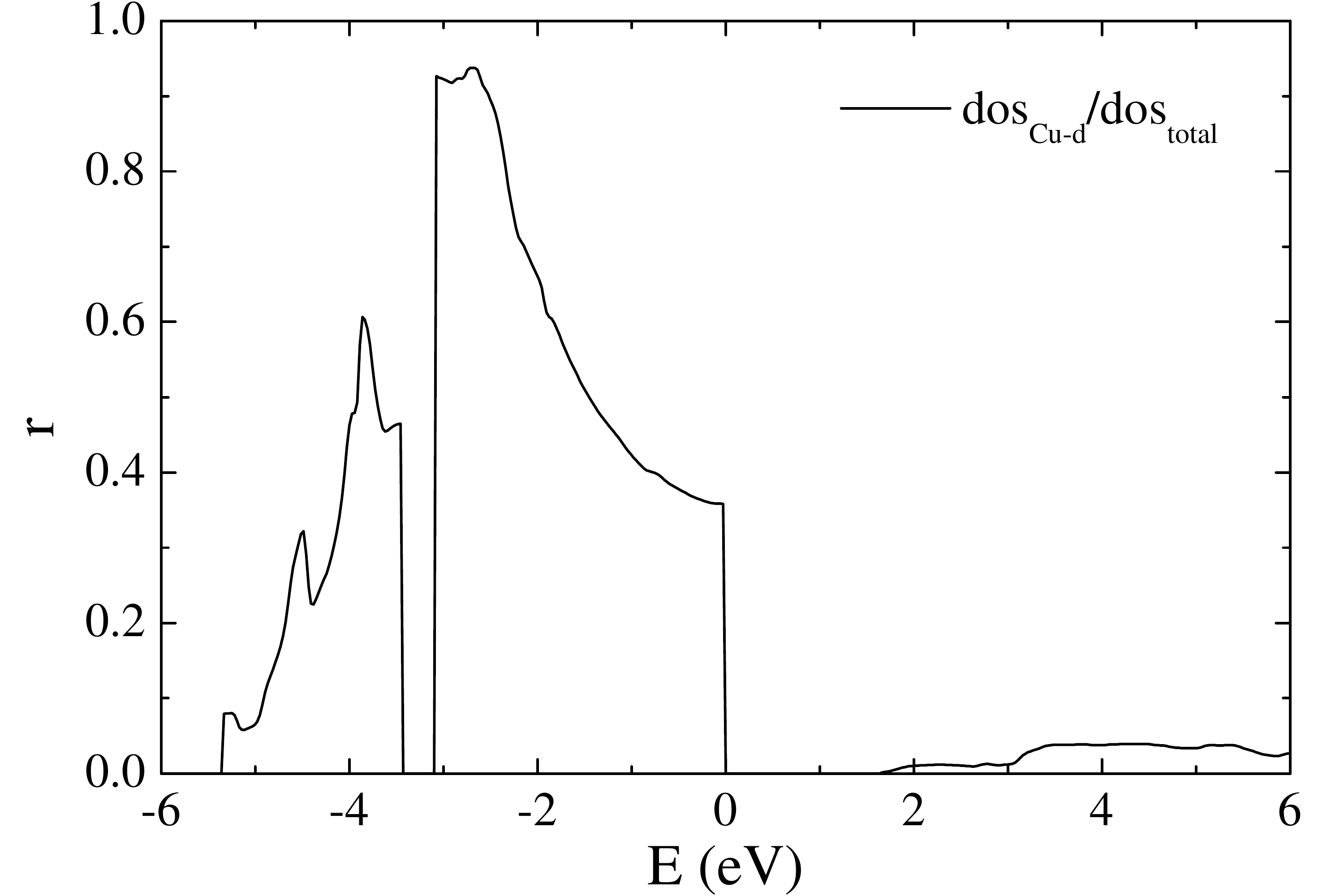}
\caption{Ratio of the Cu $d$ contribution to the total density of states,
as in Fig. \ref{dosfull}.}
\label{dosratio}
\end{figure}

The second issue that is important for optical calculations
is the band gap.
Calculation of the band gap in CuI is complicated by two issues, namely,
self-interaction errors associated with the localized Cu $d$ electrons 
and correct positioning of the Cu $s$ derived conduction bands with
respect to the I $p$ derived valence bands. Because of this combination
the band gap is underestimated even using hybrid functionals and
full self-consistent GW calculations. \cite{pishtshev}
This is also the case for our GGA+U calculations,
as seen in Fig. \ref{dosfull}, which shows the density of states over
a wider energy range.
Fig. \ref{dosratio} shows the relative contribution of the Cu $d$ states
to the the electronic structure.
One could in principle shift the band gap by applying fitted U parameters
to other orbitals. However, while it is physically clear that there is
a need to correct the binding of the localized Cu $d$ levels, related to
self interaction errors, \cite{sic}
there is no clear physical basis for applying
U parameters to other orbitals in CuI, and in fact the remaining error
is more likely associated with the density functional band gap underestimation
associated with the exchange correlation potential discontinuity.
\cite{perdew}
Here we relied on the experimental direct gap (3.1 eV),
\cite{I-CuI_1,CuI_gap2}
and applied a rigid shift of the conduction bands to match it.

\begin{figure}[tbp]
\includegraphics[width=\columnwidth]{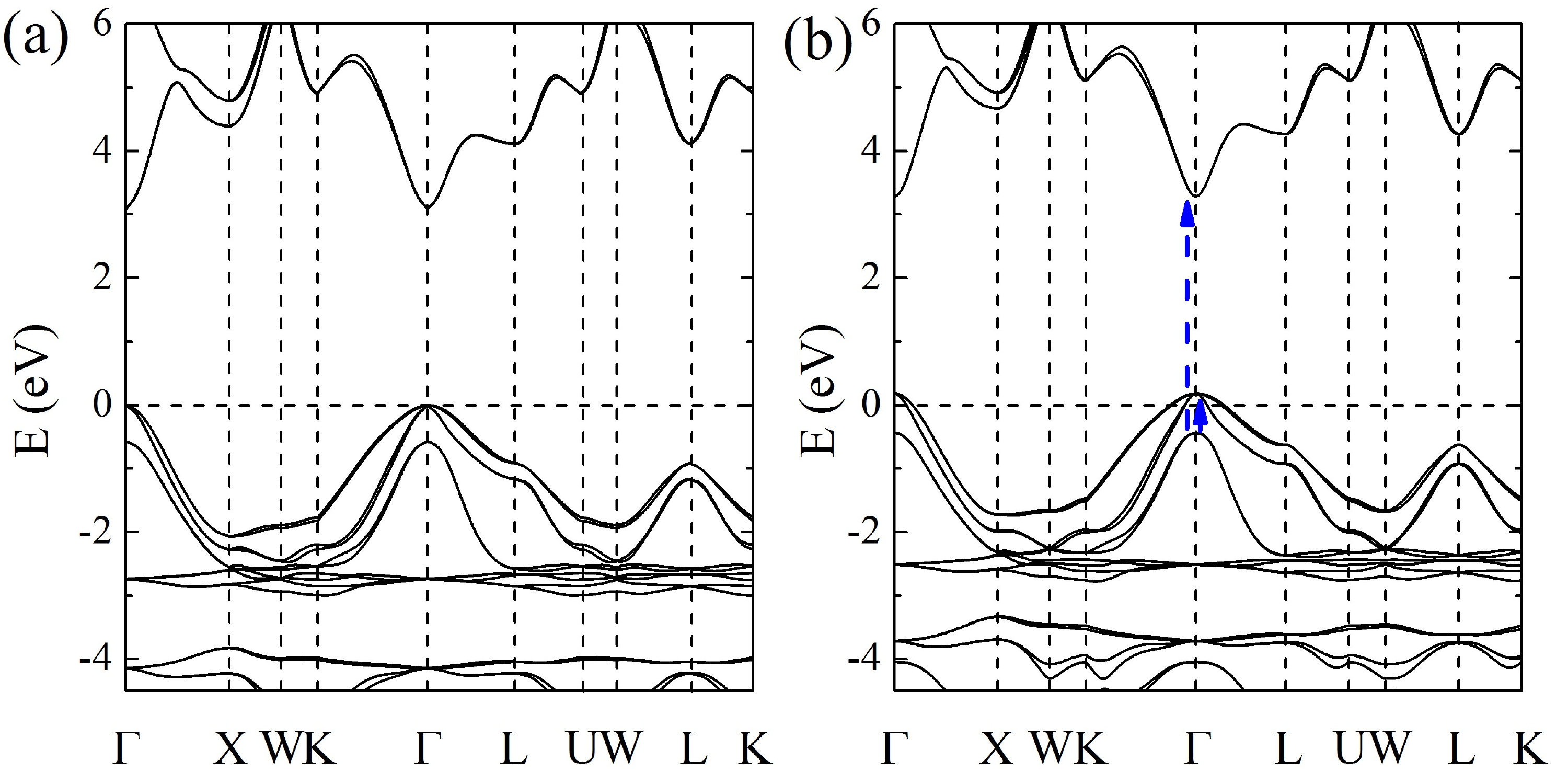}
\caption{The valence band structure of CuI as obtained from
hybrid functional calculations (left) and the band structure
GGA+U band structure used in the optical and transport calculations (right).
Optical transitions are marked with arrows.
The solid arrow indicates optical transitions in infrared and
the dashed arrow indicates optical transitions in ultraviolet.}
\label{band}
\end{figure}

We then modeled  $p$-type CuI using the virtual crystal approximation.
The virtual crystal approximation is an average potential approximation.
It goes beyond rigid bands and specifically includes composition dependent
distortions of the band structure, although in the present case
these distortions are weak.
Fig. \ref{band} shows
the band structure from the hybrid functional calculations, with the
shift to reproduce the experimental gap, compared with the virtual
crystal band structure at 0.025 holes per formula unit.
Optical properties and plasma frequencies were calculated using the optical
package of WIEN2k, using virtual crystal band structures, obtained for each
doping level.
In this code, the plasma frequency tensor is calculated as an integral over
the Fermi surface of the square band velocity (related to diagonal momentum
matrix elements). \cite{draxl}
We also did rigid band calculations for the transport
related properties using BoltzTraP. This calculation differs from
the plasma frequency calculation in that it does not include distortions
of the band structure from changes in the electron count.
These distortions are
included in the virtual crystal approximation, but not in the rigid band
approximation.
Furthermore BoltzTraP uses an analytic expression for the
band velocity based on a very fine grid Fourier interpolation of the energy
eigenvalues, different from the optical package of WIEN2k.
We find that for CuI the rigid transport calculations yield similar results to
the virtual crystal
approximation for the plasma frequency.

\section{Results and Discussion}

The band structure (Fig. \ref{band}, shown for 0.025 holes per Cu, i.e.
$p$=4.5x10$^{20}$ cm$^{-3}$ in the right panel),
illustrates one of the main conundrums in TC materials:
a TC must be both transparent and conducting at the same
time under the same conditions. The band gap of CuI, as is known
from experiment, is 3.1 eV, meaning that intrinsic CuI of sufficient
quality should be transparent for almost all of the visible spectrum.
It is also known from experiment and expected from the chemistry of compounds
of monovalent Cu that the compound naturally forms $p$-type and can
be relatively easily $p$-type doped, and then becomes conducting. The band
structure shows dispersive bands at the valence band maximum (quantified
below in terms of a transport effective mass)
consistent with reasonable conduction.

The dispersive nature of the bands arises because, as seen in the density
of states, they have antibonding Cu $d$ - I $p$ character, similar to
other $d^{10}$ semiconductors. \cite{williamson}
This is clearly seen in the density of states, and in particular the
Cu $d$ contribution (Fig. \ref{dosratio}). The antibonding I $p$ - Cu $d$
nature of the states at the top of the valence band causes them to be
higher in energy that a pure I $p$ valence band would be, leading to
increased disperion as has been discussed in other compounds.
\cite{williamson}

The problem is that conduction requires doping.
CuI, like many materials shows several bands near the valence
band maximum, including the energy range from -3.25 eV to -1.65 eV, which
corresponds to the visible. $p$-type doping introduces empty states at
the valence band maximum, and then one may anticipate transitions from
the deeper bands to the empty states, with associated absorption of visible
light.
In known high performance $n$-type TCOs, e.g.
In$_2$O$_3$, BaSnO$_3$ and ZnO, there are no conduction bands
with energies that
would allow direct transitions in the visible from the conduction band
minimum. \cite{mryasov,singh-1991,janotti}
Thus, while undoped CuI is expected to transparent due to the
band gap, and
doped CuI is expected to conduct due to the dispersion of the
bands at the valence band maximum, it may at first sight be
unclear how doped CuI can retain transparency.

\begin{figure}[tbp]
\includegraphics[width=0.9\columnwidth]{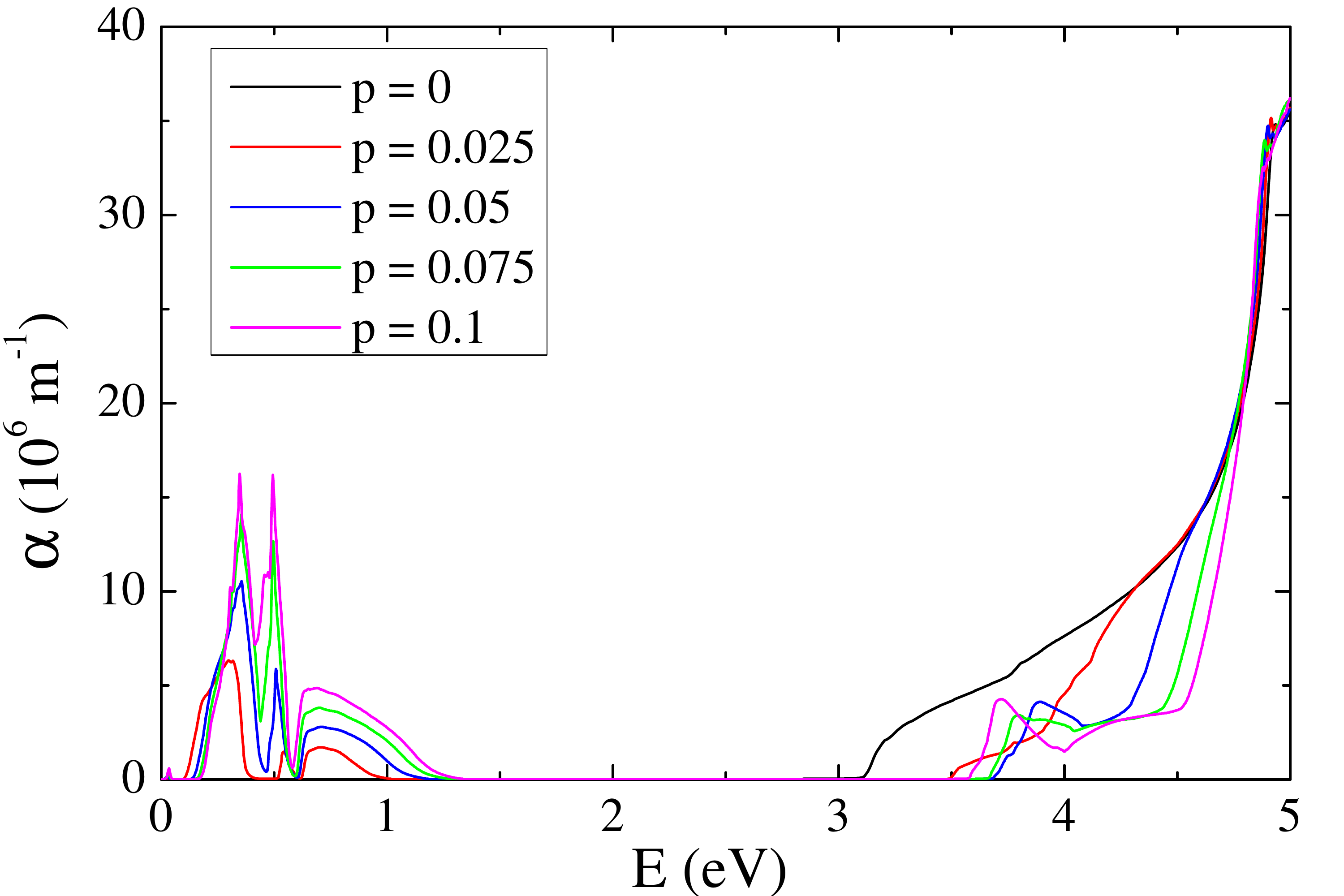}
\caption{Calculated absorption coefficient due to
interband transitions for $p$-type
CuI as a function of doping level in carriers per Cu atom.
In terms of carrier concentration,
0.1 holes per Cu correspond to $p$ = 1.80 x 10$^{21}$ cm$^{-3}$. In
addition there will be a Drude part in the infrared with a width
that depends on the scattering.}
\label{absorb}
\end{figure}

We begin with the optical properties of $p$-type CuI, based on
virtual crystal calculations that explicitly include doping
induced empty states at the top of the valence bands and transitions
into these states.
Fig. \ref{absorb} shows the absorption spectra for different
$p$-type doping levels based on interband transitions.
For zero doping there is a strong absorption edge at the direct band
gap due to dipole allowed transitions from the valence band maximum
to the conduction band minimum. These transitions are removed by $p$-type
doping as the initial states become unoccupied. This is basically
a Burstein-Moss shift. The strong absorption feature above the
gap comes from the second split off hole band for heavily doped material,
as shown, while additional strong absorption in the infrared appears
from interband transitions within the valence bands.
In addition to the interband transitions,
there will be a Drude component, which
depends on the doping dependent plasma frequency (see below) and
the usually sample dependent
scattering rate. This will yield additional infrared absorption, which
we do not include.

As mentioned,
undoped CuI is a semiconductor with a direct band gap and onset of
absorption at $>$ 3.1 eV.
Doping introduces additional strong absorption peaks
in the infrared ($<$1.3 eV) due to interband transitions.
This is in contrast to many $n$-type TCOs, such as ITO,
ZnO, and BaSnO$_3$,  and reflects the availability of
valence bands to participate in transitions to empty states at the
valence band maximum.
Thus, $p$-type CuI is not transparent for infrared light.

The key present finding of our study as regards optical properties
is that there are no significant interband transitions
in the visible part of the spectrum for $p$-type CuI.
The origin is in the gap at $\Gamma$ between the upper $d$ crystal field
level and the valence band maximum, as well as weak matrix
elements between the $d$ states and the top valence band near $\Gamma$
(note that the symmetry of the four fold degenerate state at the
valence band maximum is the same as the top Cu d band at $\Gamma$).
This is qualitatively
similar to what was recently found for another recently discovered
$p$-type TCO, K$_x$Ba$_{1-x}$SnO$_3$. \cite{pBaSnO3_1,pBaSnO3_2}
The TC behavior of $p$-type CuI depends on this feature of
its band structure.

\begin{figure}[tbp]
\includegraphics[width=0.9\columnwidth]{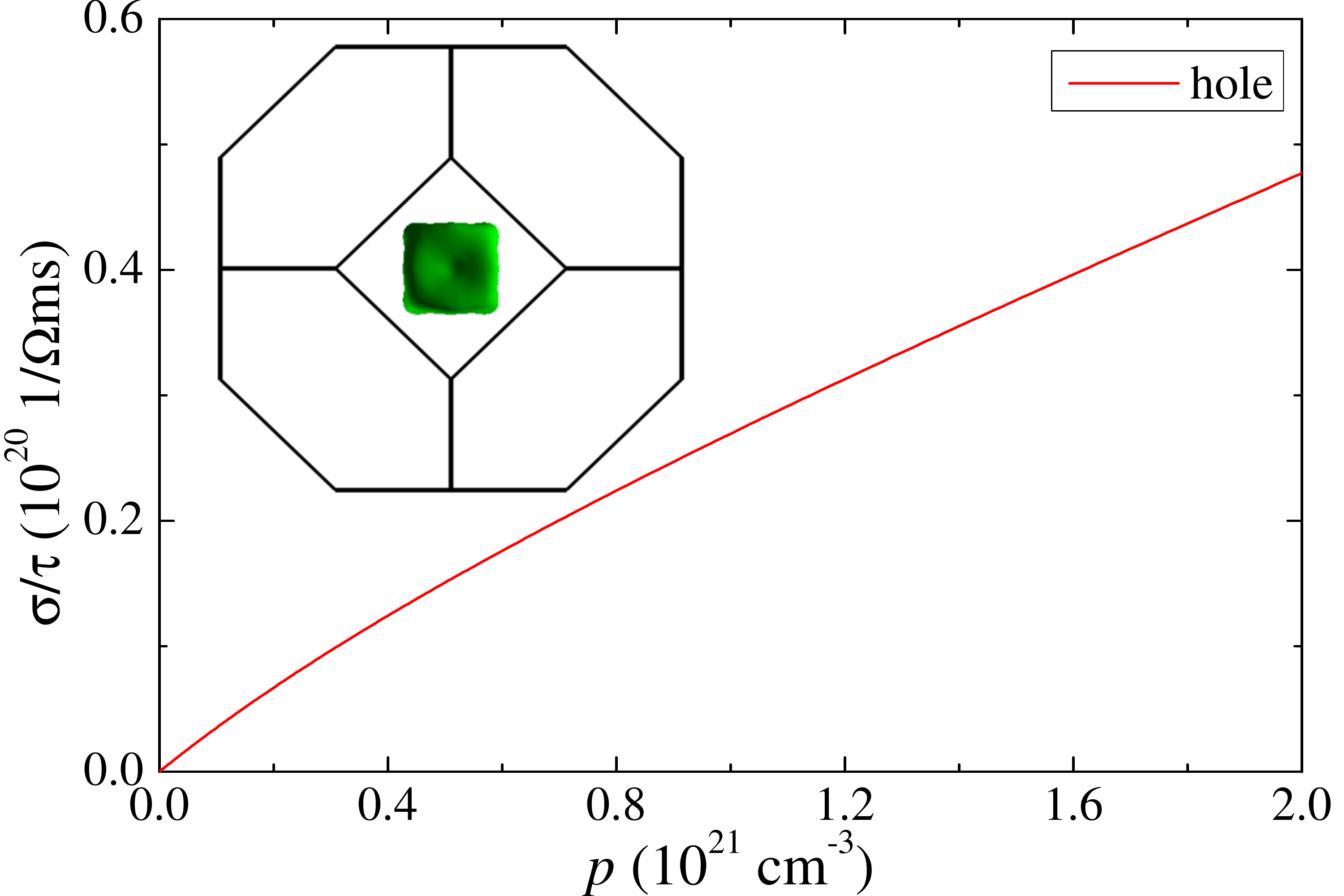}
\caption{$\sigma$/$\tau$ of $p$-type CuI (red) as a function
of carrier concentration as obtained at 300 K.
The inset is the energy isosurface for 0.025 holes per Cu,
as shown in Fig. \ref{band}.
This corresponds to a carrier concentration,
$p$ = 4.5 $\times$ 10$^{20}$ cm$^{-3}$.}
\label{p-conduct}
\end{figure}

\begin{figure}[tbp]
\includegraphics[width=\columnwidth]{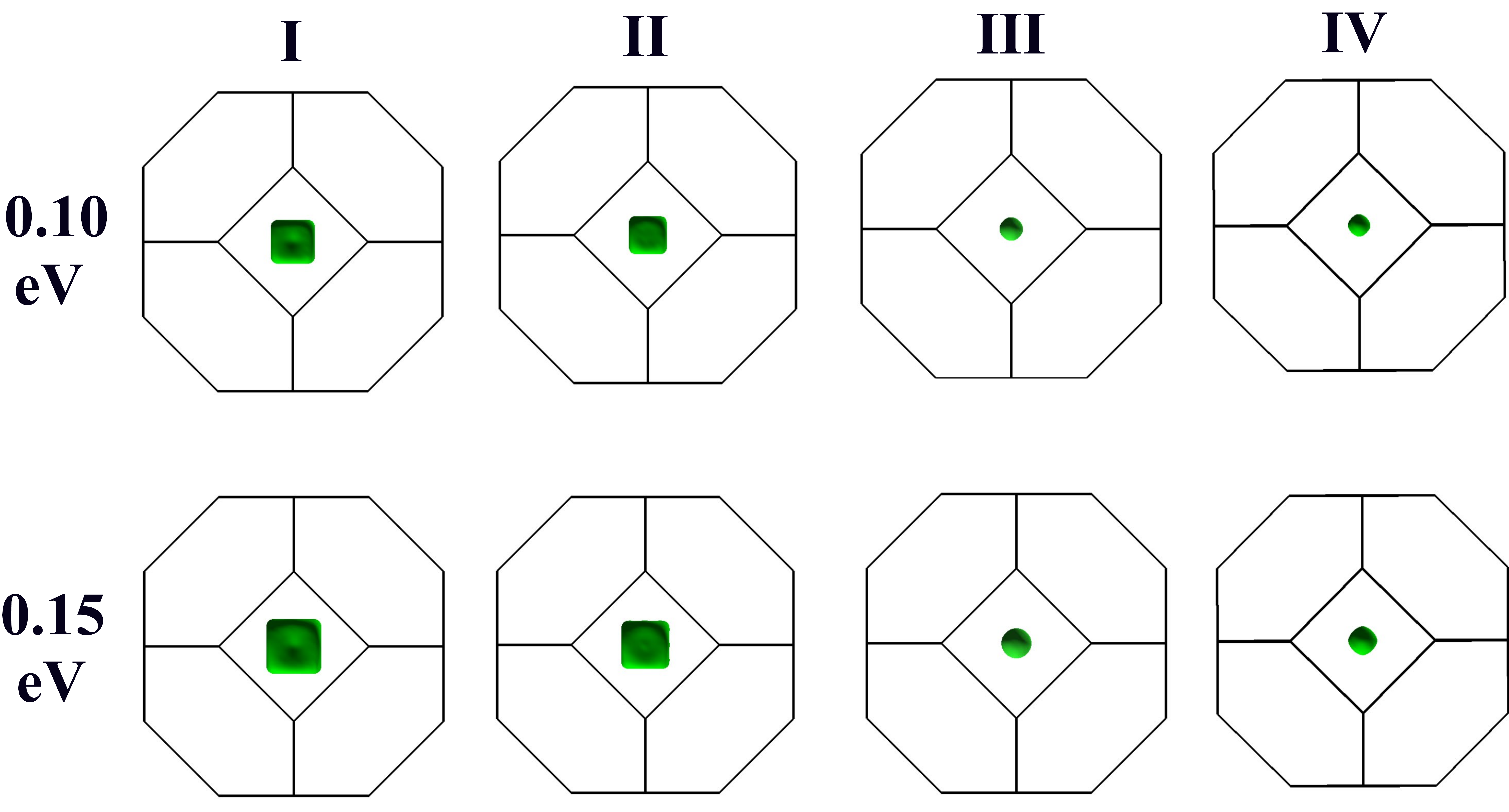}
\caption{Isosurfaces of the four top bands comprising the spin orbit split
heavy and light hole bands for different Fermi energies as shown.}
\label{isosurface}
\end{figure}

Fig. \ref{band} shows the band structure with $p$-type CuI, comparing the
hybrid functional and PBE+U calculations (both with the conduction
band shift to reproduce the experimental band gap).
The virtual crystal band structure shown in the right panel is at
a carrier concentration
of 0.025 holes per Cu ($p$ = 4.5 $\times$ 10$^{20}$ cm$^{-3}$).
The solid vertical arrows in the band structure plot
indicate the main transitions in the infrared,
which are between bands below the valence band maximum and the top band.
The dashed arrow indicates the transition between valence bands
and conduction bands with energy $>$ 3.5 eV as seen in Fig. \ref{absorb}. 

The valence band maximum at $\Gamma$ is four fold degenerate (including spin),
as is common to most zinc-blende semiconductors.
The fact that I ($Z$=53) is a heavy p-electron element, combined
with the fact that the crystal structure is non-centrosymmetric,
with significantly different atoms on the two sites (Cu and I),  results in
tiny but non-negligible spin-orbit splittings of these bands as one
moves away from $\Gamma$, so that the top four bands are formally distinct,
though neglecting these spin splittings away from $\Gamma$ they are
simply the heavy hole and light hole band characteristic of 
zinc-blende semiconductors such as GaAs.

There is an additional hole band starting $\sim$0.6 eV below the
valence band maximum. This two fold degenerate (including spin) band
is split from the valence band maximum by spin orbit. It is similar
to the corresponding band in GaAs, but the splitting is larger due
to the higher atomic number of I, relative to As. The splitting of
0.6 eV is comparable to that found in GaSb
($\sim$ 0.7 eV), containing Sb ($Z$=51). \cite{madelung,gmitra}
This band ordering is in accord with experimental data.
\cite{madelung,goldmann}
Importantly, transitions from this lower split off band
to the valence band maximum are
dipole allowed in the non-centrosymmetric zinc-blende structure. The
symmetry breaking from diamond structure is strong in CuI
because of the very different chemical natures of Cu and I.

The other key property of a good TC is high conductivity.
In practice, doping is essential for increasing the
conductivity of TCO thin films,
but usually heavy carrier concentration reduces the transparency.
Experimentally, in samples produced so far,
the transmittance of CuI is reported to reach 72\%
at a conductivity level of 280 Scm$^{-1}$.\cite{CuI_1}
Here we use the quantity $\sigma$/$\tau$ ($\sigma$ is conductivity,
and $\tau$ is an effective scattering time) to characterize the
transport properties of $p$-type CuI.
This quantity is calculated directly from the band structure
using the linearized Boltzmann transport theory with the
relaxation time approximation, as implemented in the BoltzTraP code.
Fig. \ref{p-conduct} shows this quantity as a function of doping level,
based on the rigid band approximation as determined using the
BoltzTraP code.
As seen, $\sigma/\tau$ increases roughly linearly with carrier
concentration (characteristic behavior of a parabolic band
semiconductor) at low carrier concentration.
In comparing different materials it is conventional to map the
properties onto a single parabolic band model, even in cases
such as CuI where there are multiple bands at the valence band maximum.
This allows, for example, comparison of an effective mass between different
compounds, with low effective mass indicating likelihood of good mobility
and conductivity when doped.
\cite{Ba2BiTaO6}
Matching the calculated $\sigma/\tau$
to a single parabolic band model, we obtain a
transport effective mass of $m_h$ = 0.77 $m_e$
($m_e$ is the mass of the electron) for carrier concentrations $p$
up to the $10^{19}$ cm$^{-3}$ level, and then increasing weakly to
1.05 $m_e$ for $p=10^{21}$ cm$^{-3}$. This carrier concentration
dependence shows the non-parabolicity of the bands. Non-parabolicity
with increasing mass away from $\Gamma$ point valence band maxima, especially
for the light hole band
is a characteristic of zinc-blende semiconductors consistent
with what we find, e.g. along $\Gamma$-$L$. \cite{kane}

\begin{figure}[tbp]
\includegraphics[width=0.95\columnwidth]{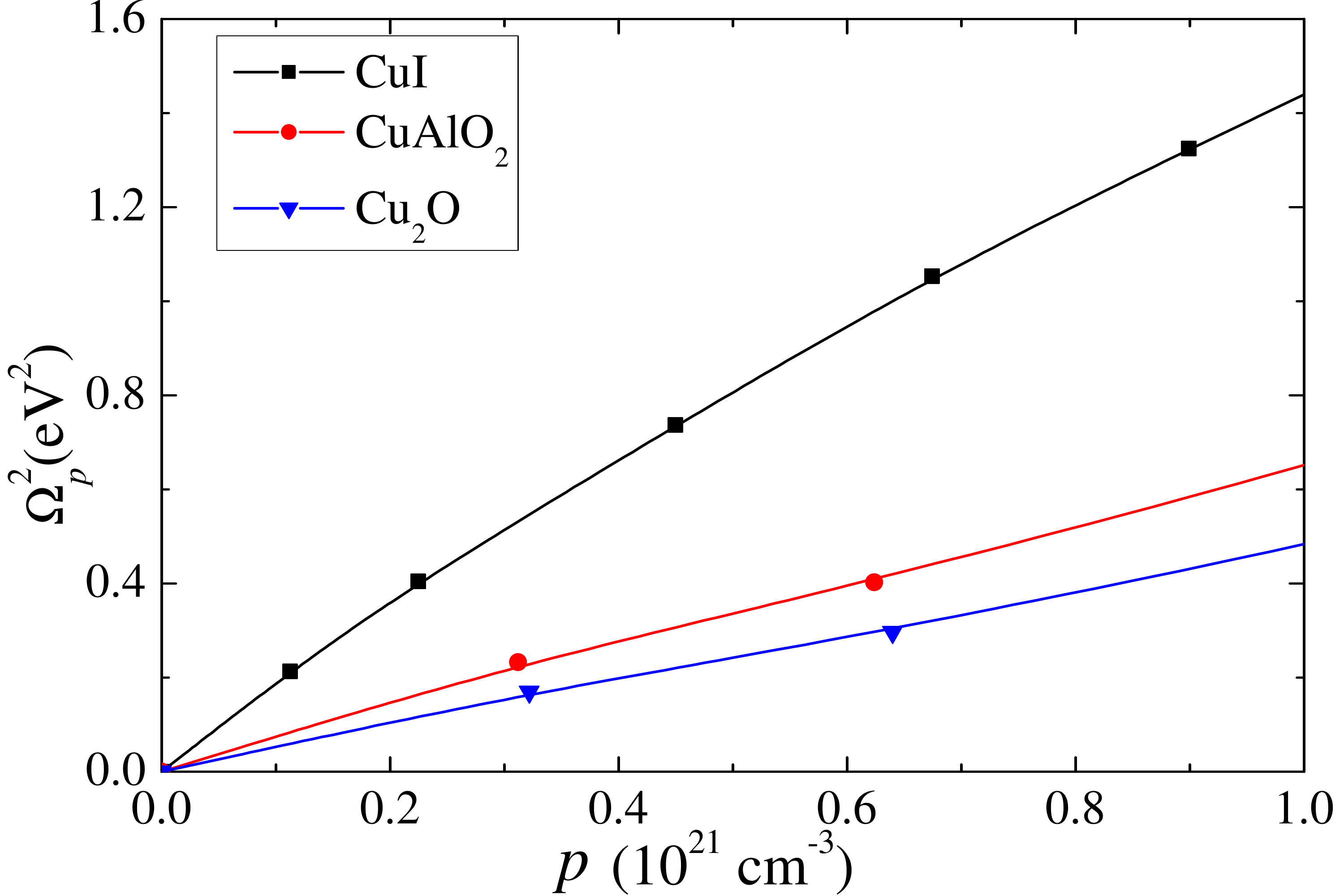}
\caption{Squared plasma frequency of $p$-type CuI (black),
CuAlO$_2$ (red) and Cu$_2$O (blue)
as a function of carrier concentration,
as obtained from virtual crystal calculations
for Cu$_{1-x}$I and for CuAlO$_2$ and Cu$_2$O
with virtual crystal $p$-type doping on the O site.}
\label{plasma}
\end{figure}

As mentioned, while it is useful to compare materials based on a transport
effective mass, the valence band structure of $\gamma$-CuI does not
show a single parabolic band at the band edge, and even considering a two
band model with a light and heavy hole band, one may see substantial
non-parabolicity.
We show the isoenergy surface of the top band
for a $p$-type doping of 0.025 holes per Cu in the inset of
Fig. \ref{p-conduct}.
Carrier pockets for the
top four bands, comprising the spin orbit split heavy and light hole bands
are shown for different Fermi levels in Fig. \ref{isosurface}.
For isolated non-interacting parabolic bands
with cubic symmetry these would be a simple spheres.
Clearly the shapes deviate strongly from spheres particularly
for the heavy hole bands. This is a consequence
of the momentum dependent splitting of the bands at the valence
band maximum as one moves away from the $\Gamma$-point.
The result of this type of complexity is generally a decoupling of the
transport and density of states effective masses in a way that
lowers the transport effective mass, and is therefore favorable for
the mobility. \cite{mecholsky,xing}
It is also to be noted that the two light hole bands have more spherical
isosurfaces. The average mass of these two light hole bands is 0.25 $m_e$,
and actually varies between 0.23 $m_e$ and 0.28 $m_e$, depending
on direction, again reflecting the anisotropy of the carrier pockets.

The calculated 0 K plasma frequencies,
$\Omega _p = \hbar\omega_p$, as a function of doping level,
are given in Fig. \ref{plasma}
along with those of CuAlO$_2$ and Cu$_2$O.
These were obtained using the virtual crystal approximation, done
by lowering the atomic number of Cu to model $p$-type doping due to
Cu vacancies, known to be important sources of $p$-type conduction
in those materials.
Conductivity in metals and degenerately doped semiconductors
depends on the plasma frequency,
$\sigma \propto \Omega ^2 _p \tau$,
where $\tau$ is an effective inverse scattering rate
(at 0 K, $\varepsilon_0\omega_p=\sigma/\tau$).
As shown in Fig. \ref{plasma}, the plasma frequency for CuI is significantly
higher than CuAlO$_2$, consistent with good TC performance, and perhaps better
performance than CuAlO$_2$ depending on the scattering rate.
It is also important to note that the doping dependence of the virtual
crystal plasma frequency and that of rigid band $\sigma/\tau$ for
CuI are very similar,
supporting a postiori the use of these approximations.

\begin{figure}[tbp]
\includegraphics[width=0.9\columnwidth]{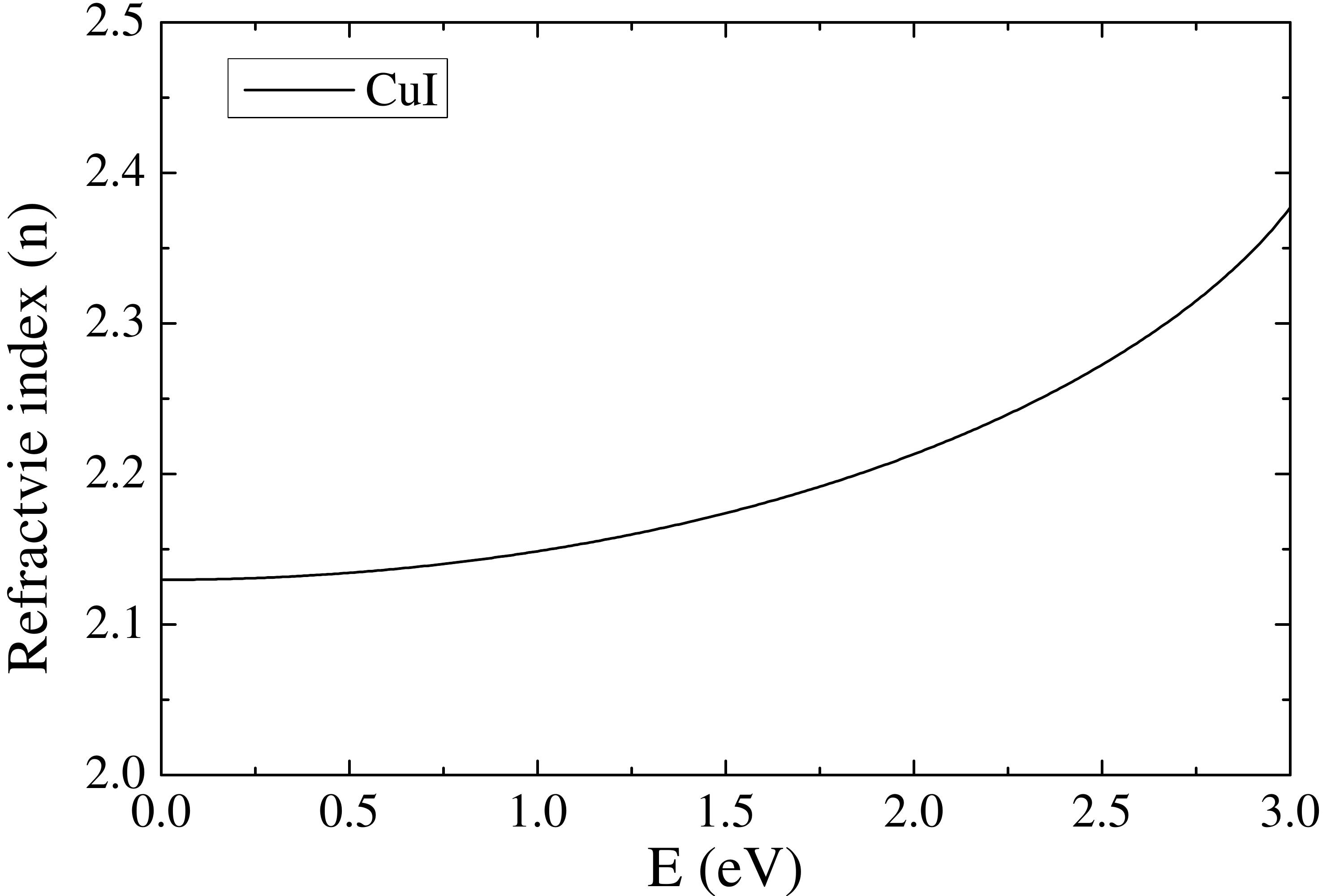}
\caption{Calculated refractive index as a function of energy.}
\label{refraction}
\end{figure}

The reported experimental
temperature dependent resistivity of heavily $p$-type degenerate
doped CuI 
is consistent with a resistivity that is dominated by defect
scattering, especially ionized impurity scattering
and grain boundary scattering rather than electron
phonon scattering. \cite{CuI_1}
As such, the question
arises as to the extent to which the scattering can be improved, e.g.
by improvements in sample quality. Some level of ionized impurity
scattering is inevitable in heavily doped bulk semiconductors as a result
of the dopants. This sets an upper limit on the conductivity.
This limit is very sensitive to the dielectric constant. \cite{shuai,sun}
In the case
where only ionized impurity scattering applies the mobility, 
$\mu$, varies as the square of the
static zero frequency dielectric constant.
\cite{debye}

Our optical calculations without doping and including
only the electronic response yield a low frequency refractive
index, $n(0)$=2.13, which is in good agreement with
literature values, $n(0)$=2.14 (Ref. \onlinecite{ageev}) and
$n(0)$=2.20 (Ref. \onlinecite{potts}).
The energy dependence is shown in Fig. \ref{refraction}.
Our low energy value corresponds to an optical dielectric constant,
$\varepsilon(\infty)$=4.53.

Ionized impurity scattering is, however, governed by the static
dielectric constant, including the lattice and electronic parts. 
Experimentally, CuI is invariably a $p$-type doped semiconductor due to
the presence of Cu vacancies. The resulting conductivity complicates
measurements of the static dielectric constant.
Literature values range from $\varepsilon(0)$=6.5 to $\varepsilon(0)$=15.
\cite{hanson,ageev,plendl}
In addition, it should be noted that the cuprous halides, including CuI,
are near a structural phase transition, reflecting borderline stability
of the zinc-blende structure against a more ionic rock structure, as is
seen in the pressure dependent phase transitions and in plots of structure
vs. Phillips ionicity. This leads to anomalous lattice dynamics,
affecting the validity of the Lynddane-Sachs-Teller relation
that may otherwise be used to estimate $\varepsilon(0)$ from phonon
measurements as well as properties
such as thermal conductivity. \cite{martin,hennion,park,mukhopadhyay}

We calculated the lattice part of the dielectric constant and the Born
charges with the PBE GGA using VASP. We also calculated
the electronic dielectric constant in the same way. The electronic
dielectric constant from these PBE GGA calculations was 4.77, slightly
larger than from the PBE+U calculations with WIEN2k as might be anticipated
in view of the lower band gap in the PBE GGA calculations.
The difference betweem the PBE GGA and the PBE+U calculations of the
electronic dielectric constant with WIEN2k
(including for the PBE+U calculation the shift of the conduction bands
to match the experimental gap)
amounts
to $\sim$5\%, which implies that the effect of the density functional
band gap error on the electronic dielectric constant is relatively
small in CuI.

The electronic structure results show a near ionic situation, with occupied
I $p$ bands, and  states near the conduction band minumum having Cu $s$
character. As such one may write a nominally ionic model, Cu$^+$I$^-$,
with nominal charges $\pm$1 on the two ions.
We obtained calculated Born charges $Z^*=\pm$1.10, i.e. weakly enhanced
from these nominal values of $\pm$1.
Such an enhancement in the zinc-blende structure
is unusual, and presumably reflects the significant ionicity of the 
compound combined with cross gap hybridization between the I $p$ derived
orbitals in the valence bands and the Cu states in
the conduction bands, consistent with a recent discussion of the bonding
based on first principles calculations. \cite{pishtshev}
However, the enhancement of the Born charges in CuI is not large enough
to lead to a large enhancement of the dielectric constant. Our
calculated value is $\varepsilon(0)$=6.3, which is at the lower end of the
literature experimental values, supporting the direct low temperature
electrical measurements of Hanson and co-workers. \cite{hanson}
It should be noted, however, that while smaller than the dielectric
constant of high performance halide semiconductors, such as TlBr
or the Pb halide perovskite solar absorbers, \cite{du2010enhanced,ming}
it is still sufficient that the limit on the mobility due to
ionized impurity scattering is well below reported values. For
example, Grundmanm and co-workers \cite{CuI_3} report a mobility fit to the
Shockley expression for a degenerate semiconductor and obtain a constant
of $\sim$0.25 $c_s$. Here $c_s$ is the coefficient for the
scattering rate for electrons of mass
$m_e$ with dielectric constant unity. Within such models the
scattering rate is proportional to the mass and inversely proportional
to the square of the dielectric constant. Our dielectric constant of
$\sim$6 would leave room for a very substantial improvement.
While this is a very crude estimate and it ignores other
sources of scattering, including the unavoidable
electron-phonon scattering, it does still imply that substantial improvement
in the mobility remains possible from improvement in sample quality.
Furthermore, we note that the relatively modest dielectric constant
of CuI suggests that there may be room for improving the electrical
properties by for example alloying as in other
cases. \cite{williamson}
Alloying to increase the value of $\varepsilon(0)$
has discussed
and shown experimentally to be effective for another material. \cite{shuai}
We do, however, note that alloying involves alloy scattering
and increased possibilities for
defects, which may work against the mobility.

\section{Summary and Conclusions}

We present first principles calculations of optical and electronic
properties of $p$-type CuI, explaining the nature of its high
transparency and conductivity.
The calculations show that although $p$-type doping of CuI
produces strong optical absorption in the infrared below 1.3 eV,
high transparency in the visible is retained due to the specific
band structure.
We also find a band structure consistent with
reasonable conduction based on the calculated transport functions.
The high visible transparency of doped CuI is a characteristic
that depends on the details of the electronic structure and goes beyond the
often quoted requirements of low effective mass and sufficient band gap
for transparent conductors. In addition we find a dielectric constant
that while smaller than other high performance halide semiconductors is
nonetheless large enough to imply that the mobility of CuI samples reported
in literature can still be substantially improved by improvements in 
sample quality.

\acknowledgements

This work was supported by the Department of Energy through the S3TEC
Energy Frontier Research Center, Award DE-SC0001299/DE-FG02ER46577.

\bibliography{reference}

\end{document}